\documentclass[conference, nofonttune]{IEEEtran}
\IEEEoverridecommandlockouts

\usepackage[acronyms,nonumberlist,nopostdot,nomain,nogroupskip]{glossaries}
\usepackage[numbers,sort&compress]{natbib}
\usepackage{xspace}
\usepackage{dblfloatfix}

\AtBeginDocument{%
	\providecommand\BibTeX{{%
			\normalfont B\kern-0.5em{\scshape i\kern-0.25em b}\kern-0.8em\TeX}}}

\usepackage[utf8]{inputenc}
\usepackage{xcolor}
\usepackage{amsmath}
\usepackage{amssymb}
\usepackage{bm}

\usepackage[acronyms,nonumberlist,nopostdot,nomain,nogroupskip]{glossaries}
\usepackage{booktabs}
\usepackage{lipsum}
\usepackage{tabularx}
\usepackage[noend]{algorithmic}
\usepackage{algorithm, setspace}
\usepackage{array}
\usepackage[draft]{hyperref}
\usepackage{balance}

\usepackage{multirow}

\renewcommand{\figurename}{Fig.}
\usepackage[font=footnotesize]{subcaption}
\usepackage[font=footnotesize]{caption}

\usepackage{mathtools}

\newcommand{\beq}{\begin{equation}}
	\newcommand{\eeq}{\end{equation}}
\newcommand{\bit}{\begin{itemize}}
	\newcommand{\eit}{\end{itemize}}

\newacronym{6g}{6G}{sixth generation}
\newacronym{5g}{5G}{fifth generation}
\newacronym{snr}{SNR}{signal-to-noise ratio}
\newacronym{sinr}{SINR}{signal-to-interference-plus-noise ratio}
\newacronym{bw}{BW}{bandwidth}
\newacronym{mod}{Mod}{Modulation}
\newacronym[plural=\gls{cnn}s,firstplural=convolutional neural networks (CNNs)]{cnn}{CNN}{convolutional neural network}

\newacronym{iq}{I/Q}{in phase/quadrature}
\newacronym{mbc}{MBC}{modulation and bandwidth classification}
\newacronym{ml}{ML}{machine learning}
\newacronym{phy}{PHY}{physical layer}
\newacronym{cvl}{CVL}{convolutional layer}
\newacronym[plural=\gls{dnn}s,firstplural=deep neural networks (DNNs)]{dnn}{DNN}{deep neural network}
\newacronym{mmwave}{mmWave}{millimeter wave}
\newacronym{dsp}{DSP}{digital signal processing}
\newacronym{dsa}{DSA}{dynamic spectrum access}
\newacronym{ism}{ISM}{industrial, scientific and medical}
\newacronym{csi}{CSI}{channel state information}
\newacronym{fcc}{FCC}{Federal Communication Commission}
\newacronym{rfp}{RFP}{radio fingerprinting}
\newacronym{sdr}{SDR}{software-defined radio}

\newacronym{pus}{PUs}{primary users}
\newacronym{sus}{SUs}{secondary users}
\newacronym{iot}{IoT}{Internet of Things}
\newacronym{mimo}{MIMO}{multi-input, multi-output}
\newacronym{mum}{MU-MIMO}{multi-user \gls{mimo}}
\newacronym{sum}{SU-MIMO}{single-user \gls{mimo}}
\newacronym{iui}{IUI}{inter-user interference}
\newacronym{isi}{ISI}{inter-stream interference}

\newacronym{wlan}{WLAN}{Wireless LAN}
\newacronym{wlans}{WLANs}{Wireless Local Area Networks}
\newacronym{rlnc}{RLNC}{Random Linear Network Coding}
\newacronym{drx}{DRX}{Discontinuous Reception}
\newacronym{cpu}{CPU}{Central Processing Unit}
\newacronym{soc}{SoC}{system-on-chip}
\newacronym{dcm}{DCM}{distributed cooperative \gls{mimo}}
\newacronym{comp}{CoMP}{Coordinated Multi-Point}
\newacronym{ap}{AP}{access point}
\newacronym{sta}{STA}{station}
\newacronym{dl}{DL}{downlink}

\newacronym{mcs}{MCS}{modulation and coding scheme}
\newacronym{cfr}{CFR}{channel frequency response}
\newacronym{ndp}{NDP}{null data packet}
\newacronym[plural=\gls{ltf}s,firstplural=long training fields (LTFs)]{ltf}{LTF}{long training field}
\newacronym{vht}{VHT}{very high throughput}
\newacronym{ofdm}{OFDM}{orthogonal frequency-division multiplexing}
\newacronym{cfo}{CFO}{carrier frequency offset}
\newacronym{sfo}{SFO}{sampling frequency offset}
\newacronym{pdd}{PDD}{packet detection delay}
\newacronym{ppo}{PPO}{phase-locked loop offset}
\newacronym{pa}{PA}{phase ambiguity}
\newacronym{sbc}{SBC}{single board computer}

\newacronym[plural=\gls{cm}s,firstplural=confusion matrices (CMs)]{cm}{CM}{confusion matrix}

\newacronym{id}{ID}{identifier}
\newacronym{aoa}{AoA}{angle of arrival}
\newacronym{ul}{UL}{uplink}

\newacronym{svd}{SVD}{singular value decomposition}

\newacronym[plural=\gls{pdf}s,firstplural=probability density functions (PDFs)]{pdf}{PDF}{probability density function}

\newacronym{mse}{MSE}{mean-square-error}
\newacronym{mmse}{MMSE}{minimum \gls{mse}}
\newacronym{kkt}{KKT}{Karush-Kuhn-Tucker}
\newacronym{fhss}{FHSS}{frequency hopping spread spectrum}
\newacronym{dsss}{DSSS}{direct sequence spread spectrum} 
\newacronym{ew}{EW}{electronic warfare}
\newacronym{rf}{RF}{Radion Frequency}
\newacronym{dos}{DoS}{denial of service}
\newacronym{hitl}{HITL}{hardware-in-the-loop}

\usepackage{longtable}

\IEEEoverridecommandlockouts

\begin{document}

\title{Narrowband Interference Detection via Deep Learning\vspace{-1ex}}

\author{\IEEEauthorblockN{Clifton Paul Robinson*, Daniel Uvaydov*, Salvatore D'Oro*, and Tommaso Melodia*}
\IEEEauthorblockA{*Institute for the Wireless Internet of Things, Northeastern University, United States\\
Email: $\{$robinson.c, uvaydov.d, s.doro, t.melodia$\}$@northeastern.edu\vspace{-2ex}}
\thanks{The authors would like to thank the Office of Naval Research for the partial funding for this project under Other Transaction Authority (OTA) N00014-18-9-001.  Any opinions, findings, conclusions, or recommendations expressed in this paper are those of the authors and do not necessarily reflect the views of the funding agency.}
\vspace{-3pt}
}

\maketitle

\begin{abstract}

Due to the increased usage of spectrum caused by the exponential growth of wireless devices, detecting and avoiding interference has become an increasingly relevant problem to ensure uninterrupted wireless communications. In this paper, we focus our interest on detecting narrowband interference caused by signals that, despite occupying a small portion of the spectrum only, can cause significant harm to wireless systems. For example, in the case of interference with pilots and other signals that are used to equalize the effect of the channel or attain synchronization. Due to the small sizes of these signals, detection can be difficult due to their low energy footprint, while greatly impacting (or denying completely in some cases) network communications. We present a novel narrowband interference detection solution that utilizes convolutional neural networks (CNNs) to detect and locate these signals with high accuracy. To demonstrate the effectiveness of our solution, we have built a prototype that has been tested and validated on a real-world over-the-air large-scale wireless testbed. Our experimental results show that our solution is capable of detecting narrowband jamming attacks with an accuracy of up to 99\%. Moreover, it is also able to detect multiple attacks affecting several frequencies at the same time even in the case of previously unseen attack patterns. Not only can our solution achieve a detection accuracy between 92\% and 99\%, but it does so by only adding an inference latency of 0.093ms.

\end{abstract}


\glsresetall


\section{Introduction}
\label{sec:intro}

Due to continuous technological advancements, the wireless spectrum is becoming more and more crowded~\cite{iot_info_crunch}. It has been estimated that by 2025, there will be 152,200 Internet of Things (IoT) devices connecting to the internet per minute and that the number of active devices will surpass 25.4 billion in 2030~\cite{iot_info_crunch}, thus exacerbating the so-called threat of a spectrum crunch~\cite{noauthor_7_noyear}. One potential solution is utilizing narrowband signals which are designed to be as least intrusive as possible and occupy a narrow portion of a radio frequency (RF) band to utilize the spectrum better~\cite{narrowband_rf}. Thanks to these properties, narrowband signals have been favored by the Federal Communications Commission (FCC)~\cite{honey}, especially for those applications with a high number of transceivers (e.g., IoT) which could potentially congest the spectrum if they were to use other waveform designs, e.g., orthogonal frequency-division multiplexing (OFDM), commonly used by other wireless systems. 

\begin{figure}[!t]
    \centering
    \includegraphics[width=.9\columnwidth]{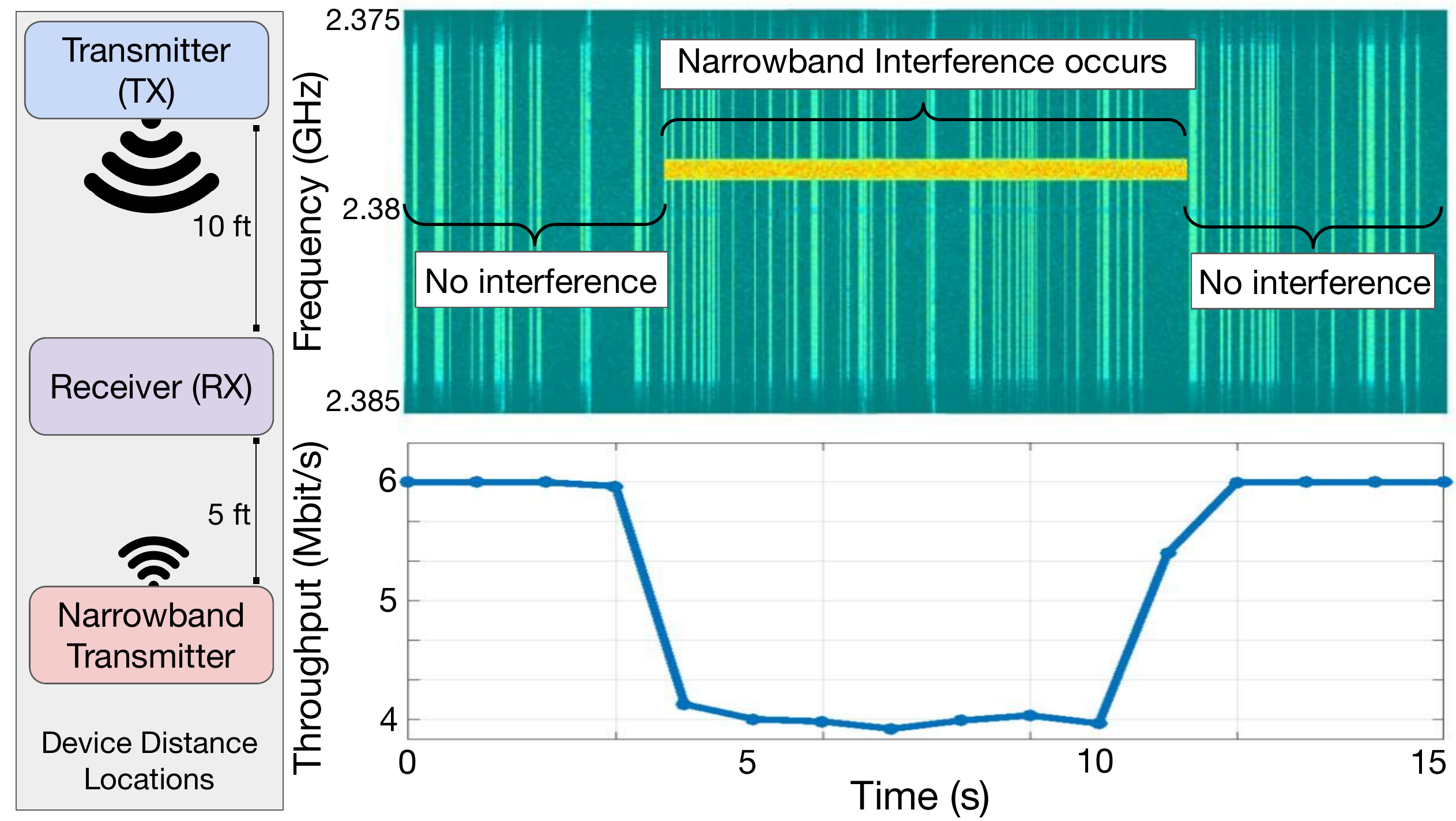}
    \caption{Narrowband interference on a 20MHz WiFi channel and the impact on interfered node's throughput, as well as the experimental setup.}
    \label{fig:jam_tp}
    \vspace{-20pt}
\end{figure}

By being small and with relatively low power, narrowband signals can indeed provide the necessary tools to mitigate the spectrum crunch. However, if used improperly, they can affect to a great extent other signals that use much larger bandwidths.

Indeed, narrowband signals can create interference (\textit{with malicious intent or not}) and have the potential to generate undesired noise and subsequent performance degradation. The most dangerous form of this type of interference is pilot jamming \cite{clancy2011efficient}, an extremely effective and energy-efficient narrowband jamming attack where a jammer targets pilots' symbols (which are commonly used by receivers to estimate and equalize the effect of the wireless channel) to substantially impair the demodulation process and potentially deny any communications.

To demonstrate how narrowband interference can severely harm a wireless system, in Fig. \ref{fig:jam_tp} we show the results of an experiment that we have conducted on the programmable over-the-air testbed Arena~\cite{bertizzolo_arena_2020}. The experiment lasts for 15 seconds. One of the radios is a WiFi node that transmits data over a 20~MHz WiFi channel. We also instantiate a narrowband transmitter that generates a 156~KHz interfering narrowband signal from seconds 4 to 11. As shown in Fig. \ref{fig:jam_tp}, although the narrowband transmitter occupies only 1.5\% of the bandwidth of the WiFi signal, it has a significant impact on the throughput of the WiFi system. Indeed, the throughput initially starts at 6~Mbit/s but drops starkly to approximately 4~Mbit/s in the case of interference. These results show that an attack occupying just 1.5\% of the legitimate signal bandwidth reduces the performance by approximately 33\%, a major reduction.

Traditional mitigation approaches against this class of interference leverage the structural properties of narrowband signals. For example, the de-facto standard countermeasure against narrowband interference is the use of spread spectrum technologies such as Code Division Multiple Access (CDMA) and frequency hopping. Although the above techniques can be extremely effective in mitigating jamming attacks~\cite{grover_jamming_2014}, there is still a need for solutions that can accurately locate ongoing interference, even if it involves just 1.5\% of the signal's bandwidth. Indeed, this knowledge is fundamental to effectively avoid interference. For example, in the case of frequency hopping, accurately locating the targeted frequency bands is necessary to determine the hopping sequence in order to avoid those bands that might be constantly under interference.

Hence, it is imperative to design mechanisms capable of detecting narrowband interference fast and with high accuracy, so as to facilitate mitigation countermeasures, which is the goal of this paper. 
Specifically, we propose a novel detection scheme to detect narrowband interference that can be implemented and utilized by leveraging IQ samples already available at the physical layer. 
We do this by utilizing deep learning (DL) techniques and specifically convolutional neural networks (CNNs) which are trained and tested with data collected over-the-air to detect different profiles of narrowband interference.

\begin{figure}[!h]
    \vspace{-10pt}
    \centering
    \includegraphics[width=.93\columnwidth]{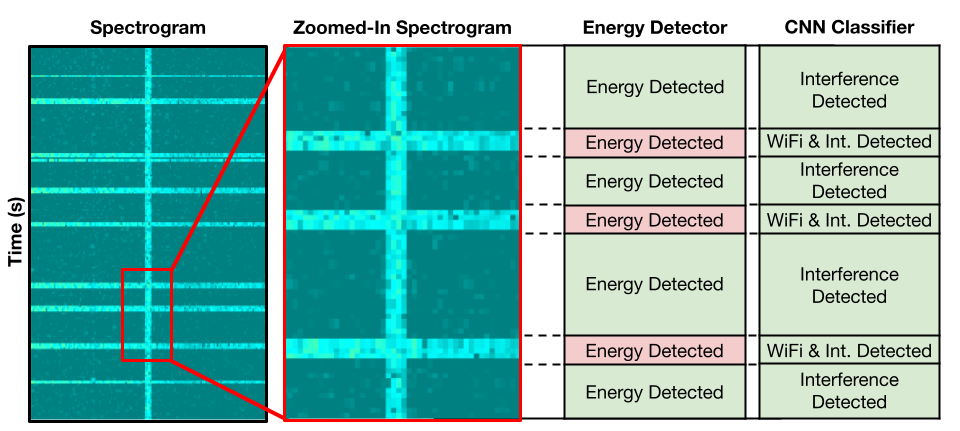}
    \caption{A comparison of an energy detector versus a CNN classifier in regards to interference detection. The energy detector can detect energy, but cannot distinguish between WiFi and narrowband interference.}
    \label{fig:energyvscnn}
    \vspace{-8pt}
\end{figure}

\textbf{Why not an energy detector?}~The first question one might ask is why we need DL for a task that might be accomplished with a simple energy detector. While energy detection is widely employed in many waveform detection problems, it lacks the necessary complexity to differentiate between the type of waveform. Take for example Fig.~\ref{fig:energyvscnn}, in this example we see a spectrogram of a waveform with a constant narrowband noise signal and three WiFi packet transmissions. With a standard energy detector, one could detect the 4 transmissions, but would fail in distinguish between narrowband interference and WiFi transmissions, hence the need for a CNN classifier able to not only detect signals but also to classify them.

We use CNNs because they offer great benefits if compared to traditional model-based solutions (generally relying upon approximations and simplifying assumptions) as they are able to learn the underlying network model directly from the data. Moreover, they can generalize to several application scenarios and perform inference in real time~\cite{restuccia_deep_2020}. 

In this paper, we make the following major contributions:
\begin{enumerate}
    \item We present a novel receiver design that embeds a CNN to provide reliable and fast detection and localization of ongoing narrowband interference. Our solution is able to characterize interference by processing received baseband IQ samples without the need to demodulate and/or decode received packets, thus achieving up to 99\% accuracy with an inference time as low as 0.093ms;
    \item We present exhaustive experimental results obtained by prototyping our solution in GNU Radio on software-defined radios (SDRs) on an over-the-air testbed~\cite{bertizzolo_arena_2020}. We assess the performance of the proposed solution for a variety of CNN architectures and configurations, identifying relevant trade-offs between complexity, latency, accuracy, and generalization of the system. Our results demonstrate the feasibility of our solution in detecting and locating narrowband interference in \textit{real time} while introducing minimal latency and overhead to the system;
    \item We show that our solution is able to detect narrowband signals over the air even in the case of previously unseen interference patterns.
\end{enumerate}



\section{Related Work}
\label{sec:relate}

DL for wireless applications has gained significant momentum in recent years from the research community~\cite{azari2022automated, d2021can, ye_power_2018, narengerile_deep_2019, ha_signal_2018, zhang_signal_2021, alhazmi_5g_2020, zhang_radio_2021, zha_deep_2019, restuccia_big_2019, restuccia_deep_2020,  punal_machine_2014}. For a recent survey on the topic, the reader can refer to~\cite{zhang_deep_2019}. Since the utilization of DL can be broadly applied to many areas within a wireless spectrum, the research takes many routes. The most common previous work falls into three categories: signal detection~\cite{ye_power_2018, narengerile_deep_2019, ha_signal_2018, zhang_signal_2021, alhazmi_5g_2020, zhang_radio_2021, zha_deep_2019, punal_machine_2014}, signal classification~\cite{zhang_signal_2021, zha_deep_2019}, and spectrum sensing~\cite{restuccia_big_2019,  DL_SS_1, DL_SS_2}.

Accurate identification of the signals in a shared spectrum is critical for both resource allocation and coexistence~\cite{zhang_deep_2019}. Due to the effectiveness in successfully detecting different types of signals, deep learning is proving to be a successful option~\cite{ye_power_2018, narengerile_deep_2019, ha_signal_2018, zhang_signal_2021, alhazmi_5g_2020, zhang_radio_2021, zha_deep_2019} even to detect complex and structured signal such as UMTS, LTE, and 5G NR~\cite{zhang_deep_2019, zhang_signal_2021}. 
In most research, we see deep learning used because it offers the generalization and versatility that traditional methods lack~\cite{zhang_radio_2021, d2021can, zha_deep_2019}. 

With spectrum sensing, the focus in the past has been divided between narrowband and wideband approaches~\cite{ss_survey}, in this paper, we only focus on the former. The majority of this research has focused on improving upon previous work by showing DL implementations offer better results compared to spread spectrum~\cite{DL_SS_1, DL_SS_2}. There is also a focus on the use of these models against previously unseen signals, showing it can still identify what is occurring when introduced to something a DL model was not trained on~\cite{DL_SS_2}. There is also research that puts a focus on latency as well as accuracy to test the real-world viability of these models~\cite{restuccia_big_2019}. Adversarial signals, specifically jamming, continues to be an essential research area due to the continuing dangers it poses to the wireless spectrum. Recently, the community has presented detection methods on wireless networks that use metrics through standard drivers and performance metrics to detect through ML~\cite{punal_machine_2014, upadhyaya_machine_2019}. In this context, we mention the use of deep reinforcement learning (DRL) to provide anti-jamming solutions~\cite{liu_pattern-aware_2019} that can avoid jamming via proper communication policies.

Different from existing work, in this paper, we tackle the problem from a radically different point of view. Specifically, we focus on designing and prototyping a device that can be used in a real wireless network and is capable of detecting and locating narrowband interference in real-time. To achieve our goal, we leverage the Arena SDR over-the-air testbed~\cite{bertizzolo_arena_2020} to perform an extensive data collection campaign. Data collected over the air is then used to train \textit{offline} a CNN that detects and precisely locates random signals, and whose testing is performed \textit{online} via over-the-air transmissions on the same testbed. We also investigate the impact of different CNN architectures on latency and accuracy and discuss how to design a system that can achieve high accuracy while supporting real-time execution with low overhead.


\section{Deep Learning for Interference Detection}
\label{sec:dl}

\begin{figure}[!h]
    \centering
    \includegraphics[width=0.92\columnwidth]{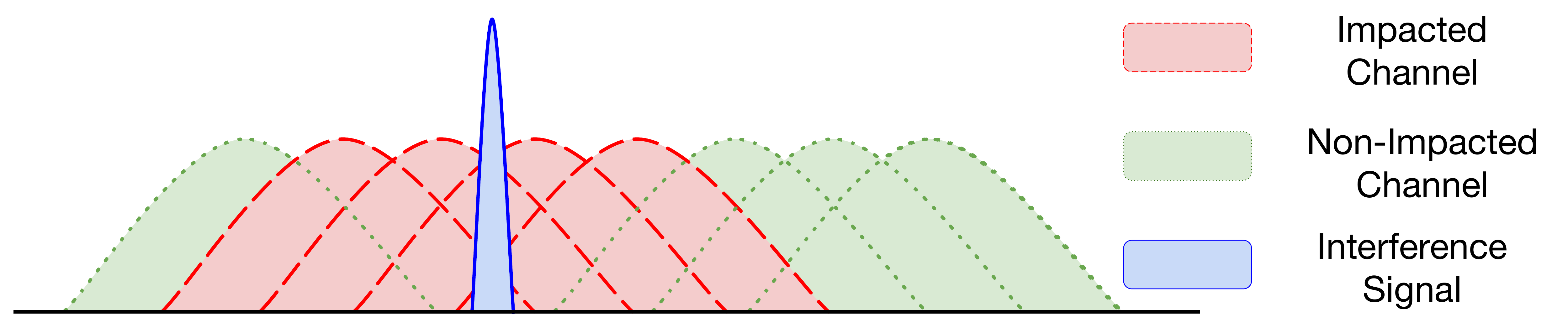}
    \caption{An example of how a narrowband interference signal can impact up to four overlapping WiFi channels at once.}
    \label{fig:narrow_sig}
    \vspace{-9pt}
\end{figure}

Narrowband signals have many important uses in the IoT thanks to their relatively low power and small spectrum usage. At the same time, these signals can still generate interference and, in those cases where they interfere (both voluntarily or not ~\cite{pirayesh_jamming_2021,clancy2011efficient}) with synchronization and equalization procedures such as pilots and reference signals, they can greatly affect network performance and can be very hard to detect~\cite{grover_jamming_2014}.
For example, Fig. \ref{fig:narrow_sig} shows how a single interfering narrowband signal can impact up to four partially overlapping WiFi channels, thus making the performance experienced by WiFi systems operating on those four channels drop.



\subsection{Proposed receiver design}
\label{sec:cnn}

\begin{figure}[!t]
    \centering
    \includegraphics[width=\columnwidth]{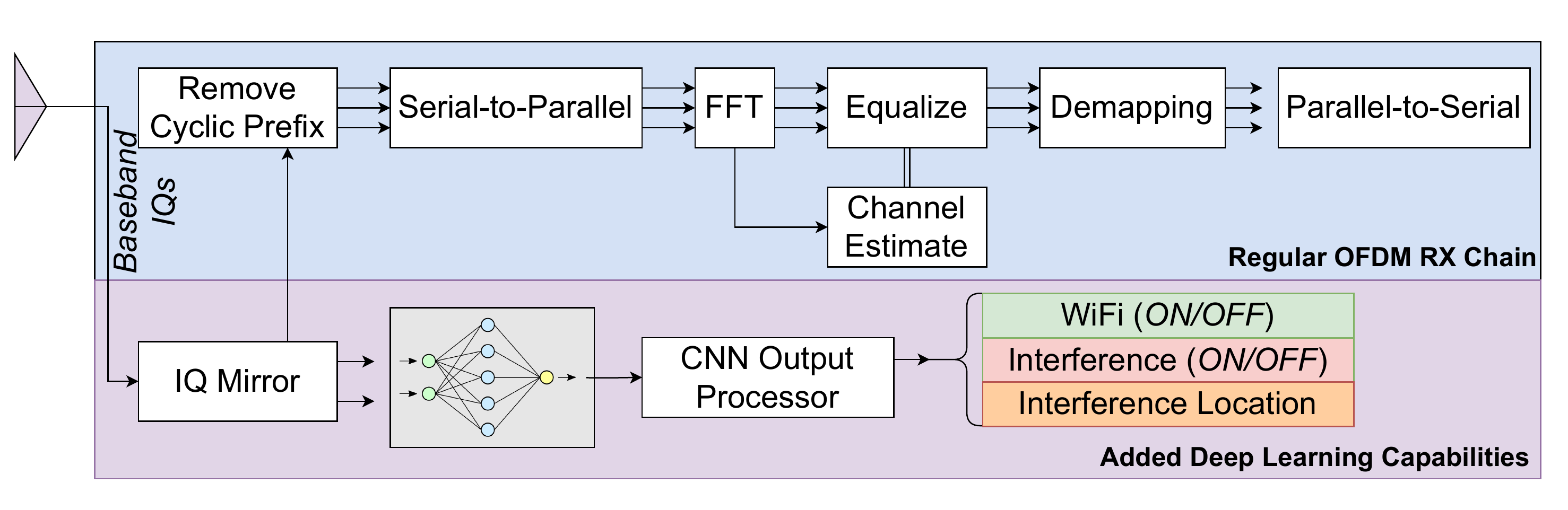}
    \vspace{-20pt}
    \caption{Receiver design block diagram and its integration with an OFDM receiver. Top: traditional OFDM receiver chain; Bottom: modules introduced by our approach and their integration with the OFDM receive chain.}
    \label{fig:rx_dl}
    \vspace{-15pt}
\end{figure}

The architecture of our proposed receiver design is depicted in Fig. \ref{fig:rx_dl}. For illustrative purposes, Fig. \ref{fig:rx_dl} shows how our solution integrates with an OFDM-based receiver, but our solution is much more general and extends to any other RF chain. Our solution is designed to introduce only minimal alterations to the receiver RF chain by adding the following two blocks: 

$\bullet$~\textbf{IQ Mirror:} it extracts the baseband IQs from the receive RF chain before they are processed. For example, in an OFDM receive chain, the IQs would be extracted before the Fast Fourier Transform block (as shown in Fig. \ref{fig:rx_dl}). This block effectively duplicates the receiver IQs and acts as a buffer. In this way, IQ mirroring makes it possible to extract the IQs without interrupting the procedures executed within the receive chain. The extraction frequency (i.e., the speed at which the IQs are stored in the buffer) is tunable so as to enable both fine- and coarse-grained waveform sampling and processing, thus making it possible to determine how many times the system samples the spectrum to detect interferences.

$\bullet$~\textbf{CNN Narrowband Interference Detector:} a block that hosts the CNN that is fed the baseband IQs to detect the location of narrowband interference. Specifically, the CNN outputs the portions of the spectrum (e.g., subcarriers) that are being affected by narrowband interference. This block will be described in detail in the following sections.


\subsection{CNN Architecture \& Training}

The CNN Narrowband Interference Detector uses baseband IQ samples from the receive chain to detect narrowband interference and must follow a design that makes it possible to be fast enough to capture even the shortest interfering signals while at the same time holding a high level of accuracy. However, this results in a trade-off between accuracy and complexity. Indeed, the smaller the input size of the model, the lower the inference latency will be because there will be much fewer computations to perform. At the same time, fewer input data can result in poor accuracy due to the lack of information to make accurate decisions. 


\begin{figure}[!b]
    \vspace{-10pt}
    \centering
    \includegraphics[width=\columnwidth]{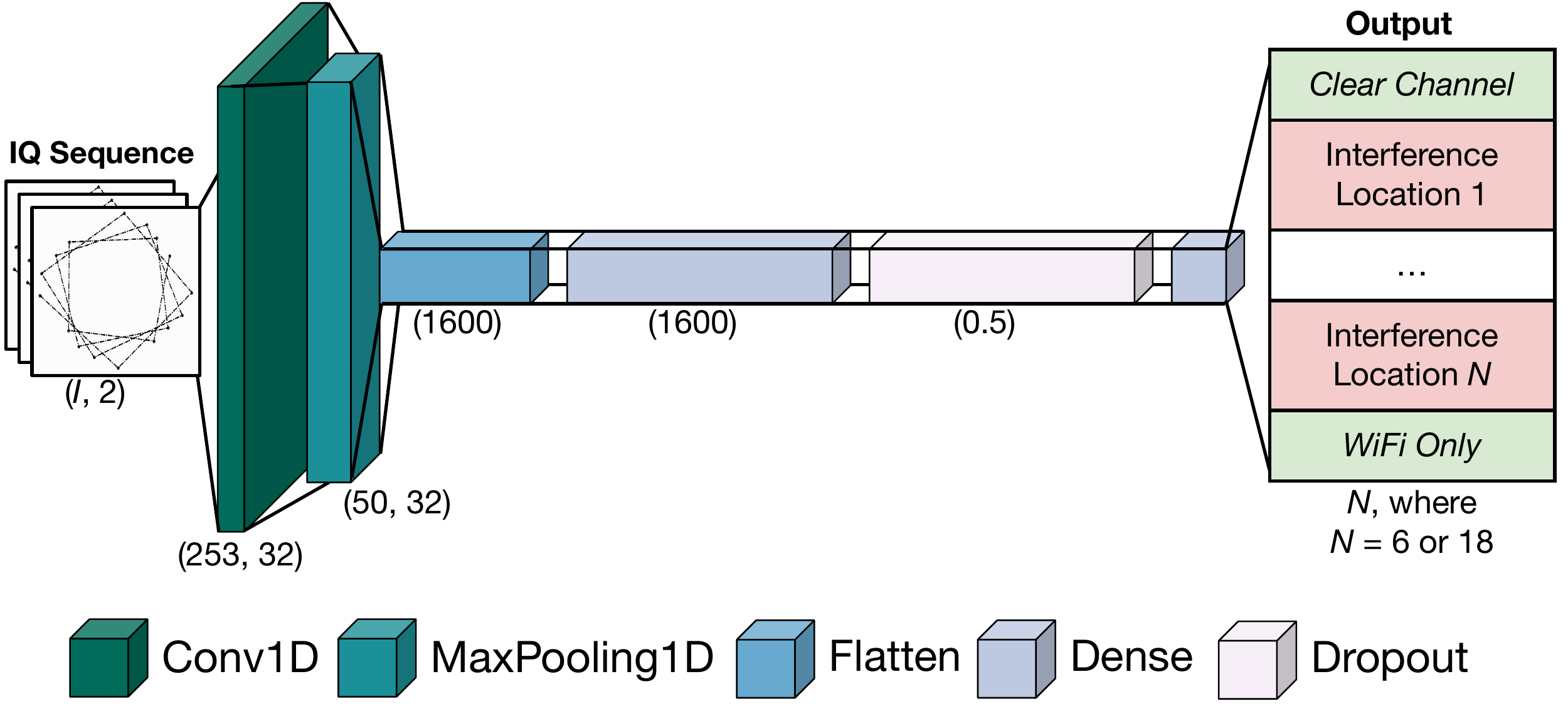}
    \caption{CNN architecture used for interference detection.}
    \label{fig:cnn_arch}
\end{figure}

To find the best trade-off, we tested multiple architectures and selected the one that delivers the best accuracy while maintaining a low inference time.
In Fig. \ref{fig:cnn_arch}, we see the architecture of the one-dimensional (1D) CNN. The input is a tensor of size ($I$, 2) representing a temporal sequence of IQs, where $I$ is the number of complex-valued IQ samples extracted from the receive chain. The impact of increasing or decreasing $I$ on model accuracy and inference latency will be investigated in Section \ref{sec:experiments}. 
The input is then processed by a single 1D convolutional (Conv1D) layer, followed by a maximum pooling (MaxPool1D) layer with filters of size 1x2 and a stride of 2. This way, the MaxPool1D helps significantly reduce the output dimension. It then goes into a flattening layer that converts the data into a 1D array to prepare it to classify the data. In Section \ref{sec:experiments}, we analyze the impact of different input sizes $I$ on the accuracy and latency of the CNN.
Then the data is sent into the first Dense layer of 1000 units and into a Dropout layer of 0.5, which cuts the number of outputs in half during active training. The Dropout layer is used to prevent the CNN from overfitting and help generalize. Finally, the data goes to the final Dense layer with Softmax activation (i.e., a logistic function that takes the outputs and normalizes them over a probability distribution) and $M = N + 2$ output neurons. In our model, $N$ represents the number of spectrum portions of interest. For example, assuming a 20MHz channel of interest, $N=4$ would split the classification domain into 4 spectrum portions (or subcarriers), each 5MHz wide.
The remaining 2 classes are used for identifying inputs containing no RF emissions (e.g., no signals) or legitimate transmissions only (e.g., WiFi only), respectively. Anytime interference is detected, the CNN will output which of those $N$ subcarriers is affected by interference. Note that $N$ is a parameter of our solution and can be used to regulate the resolution at which interference is being detected. 

\begin{figure}[!b]
    \vspace{-10pt}
    \centering
    \includegraphics[width=.95\columnwidth]{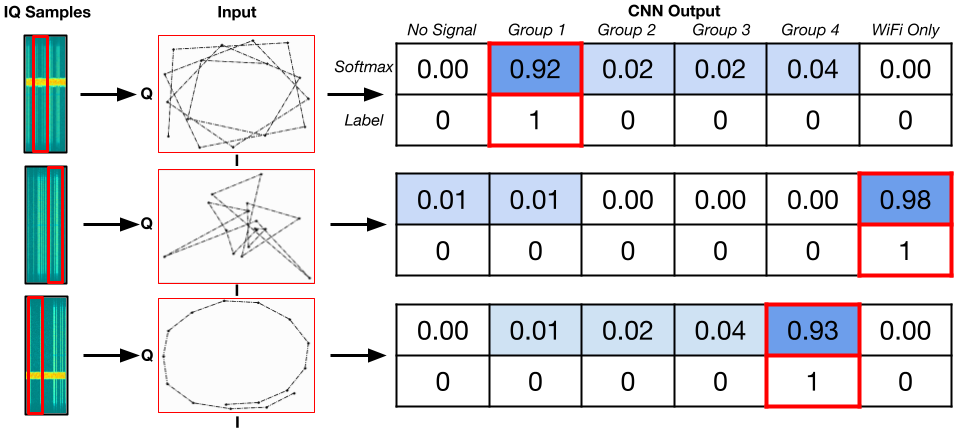}
    \caption{Inference output for several input instances in the case of $N=4$ possible interference locations.}
    \label{fig:cnn_in_out}
\end{figure}


An example with three different instances of narrowband interference is shown in Fig.~\ref{fig:cnn_in_out}. The first instance shows narrowband interference over WiFi on the first portion of the spectrum. The softmax output shows how it overwhelmingly classifies for that portion and assigns it that label. Similarly, for the middle input, only WiFi is transmitting and the classification reflects this aspect. Finally, the bottom input has interference on the fourth portion which is correctly classified by the CNN.

To train our CNNs, we use the Adam optimizer, a learning rate of 0.01, a Categorical Crossentropy loss function (CCE), and early stopping to further prevent overfitting. 

We use an 80\%, 10\%, and 10\% split to generate training, testing, and validation datasets, respectively. The datasets collected over-the-air and used to train our CNN will be described in detail in Section \ref{sec:results}. It is worth noting that while the CNN has been trained to detect one interference at a time, in Section \ref{sec:experiments} we experimentally demonstrate that this architecture can also be used to simultaneously detect multiple interfering signals on different portions of the spectrum at the same time.


\begin{figure}[!t]
    \centering
    \includegraphics[width=.9\columnwidth]{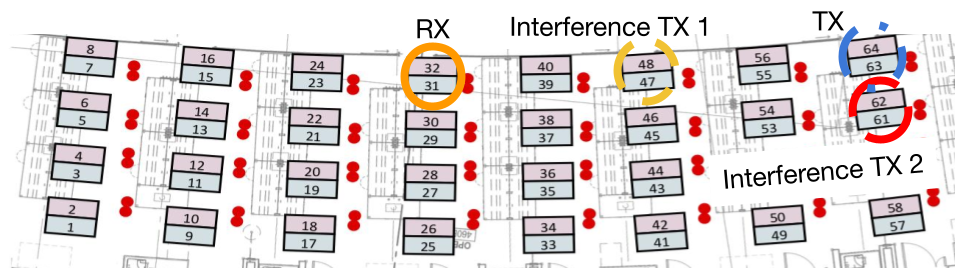}
    \caption{The layout of the Arena testbed~\cite{bertizzolo_arena_2020} with radio locations and distances.}
    \vspace{-15pt}
    \label{fig:arena}
\end{figure}

\section{Experimental Testbed Setup}
\label{sec:setup}

For our experiments and data collection, we utilize Arena, a 64-antenna SDR over-the-air testbed with support for GNU Radio ~\cite{bertizzolo_arena_2020}. Arena, shown in Fig. \ref{fig:arena}, consists of a ceiling-mounted antenna array with 12 computational servers and 24 SDRs operating in the sub-6GHz range. The testbed gives us the possibility to use a real-world environment to conduct our data collection and experiments, allowing for none of the data used to be simulated. Arena gives us also the possibility to customize employed waveforms, central frequencies, power levels, number, and location of interfering and legitimate nodes, thus offering the ideal platform to test and validate the generalizability of our solution. 

For the WiFi nodes, we use the GNU Radio implementation from \cite{bastian} and consider a bandwidth of 10MHz with 64 OFDM symbols (or subcarriers). To generate narrowband interference, we program a set of SDRs to generate narrowband gaussian noise with a bandwidth equal to 156~KHz and equal to $\frac{1}{64}^{th}$ the size of the WiFi channel. 

To test our solution, we have extended the WiFi GNU Radio receive chain ~\cite{bastian} to add both the IQ Mirror (which is implemented as a sampling block with a storage buffer) and the CNN Narrowband Interference Detector shown in Figure \ref{fig:rx_dl}. 


\section{Datasets}
\label{sec:results}

To train our CNNs, we collected data and generated two different datasets. Data is always collected over the air but on different days and with different interference profiles and distances as shown in Fig. \ref{fig:arena}.
Specifically, the first dataset includes interference signals that affect only 4 possible locations uniformly distributed into $N=4$ spectrum portions. The second dataset is more refined and includes interference signals that uniformly affect $N=16$ spectrum portions. To capture the interference location properly, we have trained two CNNs with different output classes. One, i.e., \textit{CNN4}, can with 6 output classes only capture interference on 4 spectrum portions, and determine whether or not the channel is empty or being used by legitimate nodes. Another CNN, i.e., CNN16, can instead detect interference with a higher resolution and across 16 spectrum portions. We collect two unique datasets with similar data to ensure each model trained with a dataset has similar properties and the similar data collected between the two datasets can be tested against the other model to get results for model generalization.
In both cases, we generate interference via gaussian noise that passes through a low pass filter that let us control the bandwidth and location of the narrowband interference signal.

\subsection{Dataset 1 - Interference on 4 subcarriers}
The first dataset consists of around 530,000 labeled IQ samples collected over several hours and with six different configurations (one per label): no transmissions, WiFi only, and WiFi being interfered by one narrowband signal on one of the four possible locations. To generate a balanced dataset, each label has approximately 88,000 training examples.

\subsection{Dataset 2 - Interference on 16 subcarriers}
The second dataset consists of around 1,600,000 labeled IQ samples where interference can occur uniformly across 16 locations and has a total of 18 labels. The dataset is also balanced and has 88,000 training samples per class.


\begin{figure}[t!]
    \vspace{+1pt}
    \centering
    \includegraphics[width=.95\columnwidth]{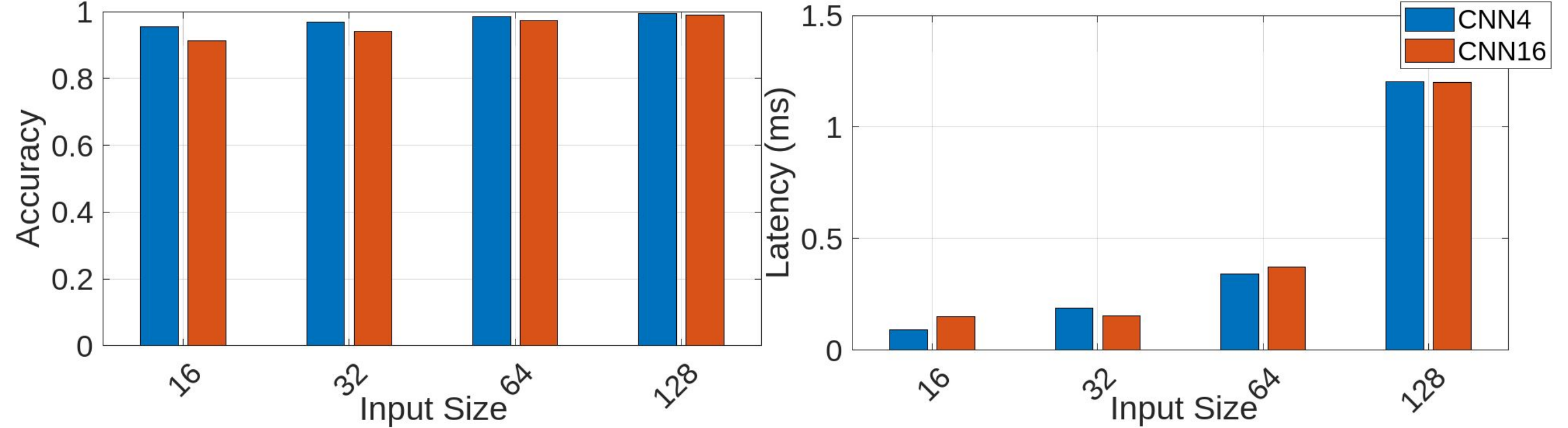}
    \caption{The accuracy (left) and inference latency (right) of CNN4 and CNN16 as a function of the input size.}
    \label{fig:cnn_acc_lat}
    \vspace{-15pt}
\end{figure}

\section{Experimental Results}
\label{sec:experiments}

In this section, we present accuracy and latency results obtained by testing our trained CNNs with online data collected over the air.

\subsection{Accuracy and inference latency} 

For both \textit{CNN4} and \textit{CNN16} we have trained different models with varying input sizes $I$ of 16, 32, 64, and 128. 

Fig. \ref{fig:cnn_acc_lat} shows the accuracy (left) and inference latency (right) of each model based on input size for both CNN4 and CNN16. 
Both CNN's perform remarkably well. CNN4's lowest accuracy is 95.5\% when $I=16$ and its highest is 99.5\% with $I=128$. CNN16 experiences a lower accuracy in general due to the higher number of classes and interference locations but can deliver 98.8\% accuracy at a higher resolution (i.e., 6 classes of CNN4 against the 18 of CNN16) when $I=128$, meaning that CNN16 can locate interference in the frequency domain with more precision while only losing 1-4\% accuracy.

\begin{figure}[!h]
    \vspace{-10pt}
    \centering
    \includegraphics[width=.9\columnwidth]{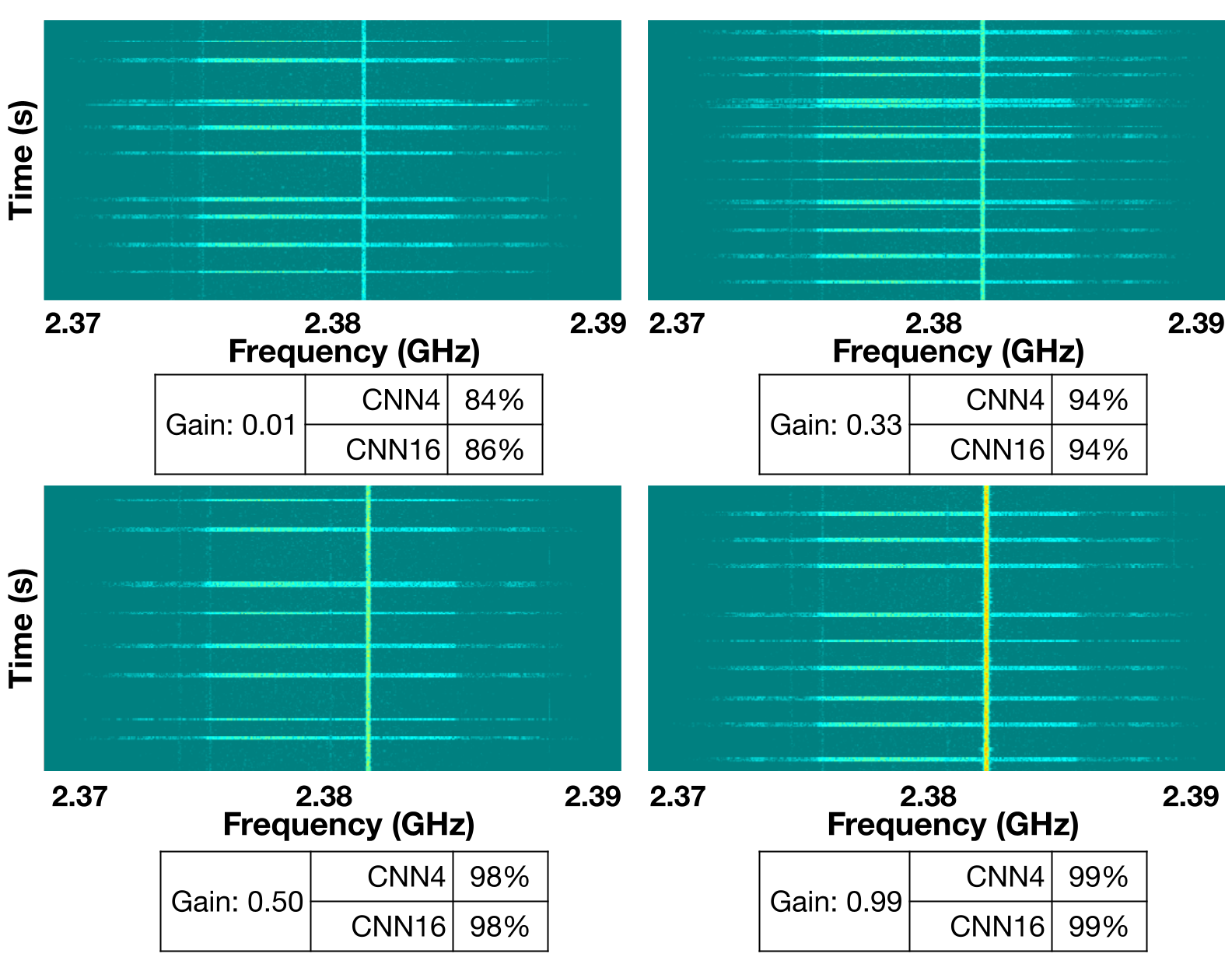}
    \caption{The different levels of gain shown in the narrowband interfering signals and the average accuracy of each CNN model per different gain value.}
    \label{fig:gain}
    \vspace{-5pt}
\end{figure}

Fig. \ref{fig:cnn_acc_lat} (right), instead, shows the inference latency of the trained CNNs as a function of the input size $I$. To measure the inference latency, both CNNs have been integrated within the GNU Radio receive chain. When $I=16$, CNN4 achieves the lowest latency value of 0.093ms, while CNN16 is able to perform inference on the IQs in 0.150ms. As expected, the inference time increases with the length $I$ of the input, and reaches 1.205ms and 1.199ms for CNN4 and CNN16, respectively. Thanks to the rather shallow architecture of both CNNs (illustrated in Fig. \ref{fig:cnn_arch}), our solution is able to deliver real-time inference and detect narrowband interference fast and with high accuracy, making it suitable for real-world applications.

\subsection{Transmission Gain Analysis}
An important aspect to demonstrate the effectiveness of AI solutions for wireless applications is to show how accurate is the AI when operating under diverse signal strength levels, which is the goal of this section. To leverage the full dynamic range of transmission power of the SDRs of Arena, in this analysis we consider a normalized gain factor in $[0,1]$ to regulate the transmission power of the narrowband interferer. In our case, we test five different gain levels  0.01, 0.33, 0.5, 0.66, and 0.99. We tested these values against CNN4 and CNN16 with two different input sizes, 16 and 128.

As seen in Fig.~\ref{fig:gain}, the narrowband signals become noticeably more potent, going from a light blue shade in the spectrogram to a darker yellow when increased. Interestingly, a gain of 0.33 results in an accuracy loss of about 1\%, while both gain values of 0.66 and 0.99 result in 99\% accuracy. As expected, a gain value of 0.01 is the one that delivers the lowest accuracy results for all CNNs. However, the accuracy loss is approximately 10\%, which still makes it possible to detect the interference despite having very low power.





\subsection{Detecting Multiple Interfering signals}

The goal of this section is to demonstrate that our solution is able to generalize across previously unseen data. Specifically, our goal is to show that our CNNs can detect multiple interfering signals being transmitted at the same time by one interferer despite these CNNs being trained to output one label only and were never exposed to multiple interfering signals at the same time. We consider both CNNs with input size $I=16$ (which is the one that delivers the lowest accuracy across our experiments, as shown in Fig. \ref{fig:cnn_acc_lat}).

\begin{figure}[h!]
    \vspace{-8pt}
    \centering
    \includegraphics[width=0.9\columnwidth]{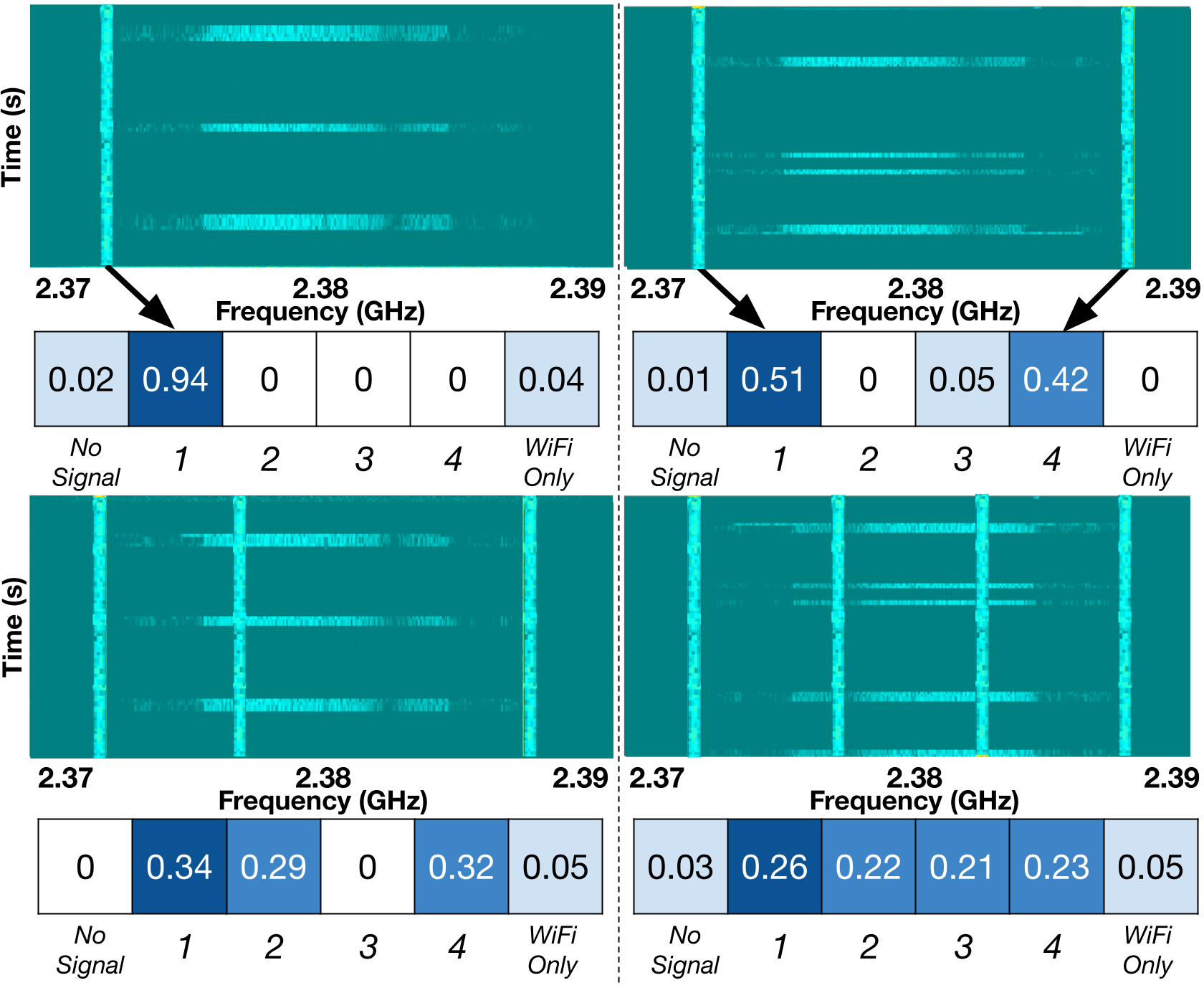}
    \caption{Detection of multiple signals using  \textit{SoftMax} outputs.
    }
    \label{fig:mult_sigs}
    \vspace{-8pt}
\end{figure}

To generalize our solution it is worth noticing that 
our CNN uses a Softmax activation function at the ultimate layer, which gives a list of probabilities that sum up to 1. Therefore, by analyzing the softmax output, multiple interfering signals can be found, as shown in Fig. \ref{fig:mult_sigs}. For example, the figure shows that despite the CNNs would output only one label, the softmax has high values for all those spectrum portions that contain narrowband interference. In the case of a single interfering signal, the softmax value for the affected subcarriers is 0.94. In the case of two interfering signals, the softmax creates an almost perfect split in detection (0.51 and 0.42). This phenomenon also occurs in the case of three and four interfering signals, where the softmax values for the affected portions of the spectrum are approximately 0.33 and 0.25, respectively. As a consequence, although the CNN is not trained to detect multiple interference signals, its softmax-based architecture can still be used to infer their existence and location.

\begin{table}[!t]
\setlength\belowcaptionskip{5pt}
    \centering
    \footnotesize
    \setlength{\tabcolsep}{2pt}
    \caption{Different types of signals tested against the CNN Models.}
    \label{tab:jam_sigs}
    \begin{tabularx}{\columnwidth}{
        >{\raggedright\arraybackslash\hsize=0.8\hsize}X 
        >{\raggedright\arraybackslash\hsize=0.8\hsize}X
        >{\raggedright\arraybackslash\hsize=0.8\hsize}X
        >{\raggedright\arraybackslash\hsize=0.8\hsize}X
        >{\raggedright\arraybackslash\hsize=0.8\hsize}X }
        \toprule
        Signal Name & Signal Type & Structure & CNN4 Acc. & CNN16 Acc. \\
        \midrule
        General Noise & Narrowband & Gaussian noise & 96\% & 91\% \\
        Packet Noise & Narrowband & OFDM Packets & 83\% & 71\%  \\
        \bottomrule
    \end{tabularx}
    \vspace{-15pt}
\end{table}

\subsection{Accuracy Against Unknown Signals}
In this section, we test the robustness of our solution against unseen interference waveforms. As mentioned earlier, our datasets contain IQ samples where narrowband interference is generated by transmitting Gaussian noise. In this section, we instead consider the case where interfering signals use an OFDM-based pulse shape similar to that used by sub-carriers on legitimate nodes. 
Both CNN4 and CNN16 have never been trained on this data or have ever seen this type of interference. Using this signal implementation, we ran experiments to test how generalized the models truly are. Both CNN4 and CNN16 use their models with an input size of 16.

As shown in Table~\ref{tab:jam_sigs}, the classification accuracy of CNN4 is equal to 96\%, which is slightly lower than the accuracy in the case of Gaussian noise (Fig. \ref{fig:cnn_acc_lat}). 
A similar drop in performance is also experienced by CNN16, which can classify and locate the new interfering waveform 
with 87\% accuracy, which is a 6\% loss in accuracy if compared to how it can classify Gaussian noise interfering signals. 
These results confirm that, at the cost of a small drop in accuracy, the proposed solution can also generalize across different waveforms that were not previously included in the training set.


\section{Conclusion}
\label{sec:end}

In this paper, we have proposed a DL-based receiver design to detect and locate narrowband interference. By utilizing CNNs, our solution is able to characterize narrowband interference starting from baseband IQ samples collected at the RF front-end. We first discussed the system design and CNN architectures and then built a prototype that we used to validate our solution with over-the-air transmissions using the Arena testbed and GNU Radio. Our results show that our solution can detect multiple interfering signals with high accuracy and within 1ms in most cases. Moreover, we have evaluated the generalization capabilities of our solution, and we have shown that it can successfully and accurately operate with previously unseen data such as multiple narrowband interfering signals with different power levels and waveform designs.


\footnotesize
\bibliographystyle{IEEEtran}
\bibliography{mybib} 
\end{document}